\documentstyle[12pt,twoside,fleqn,espcrc1,epsfig]{article}

\newcommand{\AmS}{{\protect\the\textfont2
  A\kern-.1667em\lower.5ex\hbox{M}\kern-.125emS}}

\title{Confinement in QCD}

\author{A. Di Giacomo
\address{Dipartimento di Fisica Universit\`a di Pisa
and INFN\\
        Via Buonarroti 2, 56100 Pisa, Italy}
}

\begin{document}
\maketitle

\begin{abstract}
The guiding lines of the lattice investigations on colour confinement are
reviewed, together with recent results.
\end{abstract}

\section{Introduction}
Colour is confined in nature. The experimental upper limit on abundance of
quarks $n_q$, on abundance of protons $n_p$ is
\begin{equation}
\frac{n_q}{n_p} \leq 10^{-27}\label{eq:e1}\end{equation}
corresponding to Millikan like analysis of $\sim 1 g$ of matter.
If there were no confinement a conservative expectation in the
cosmological standard model would be\cite{4}
\begin{equation}
\frac{n_q}{n_p}\simeq 10^{-12}\label{eq:e2}\end{equation} A factor
$10^{-15}$ is convincing evidence for confinement. Its smallness
also indicates that confinement cannot be explained by fine tuning
of a small parameter, but only in terms of symmetry. An analogous
situation exists in superconductivity\cite{6}. The persistence of
electric currents during years in a superconductor is explained in
terms of the change of symmetry of the ground state produced by
the Higgs phenomenon. The lifetime of the current is determined by
the characteristic correlation time of macroscopic fluctuations.

Confinement must be explained in terms of symmetry and if QCD is the theory of
hadrons it should have the relevant symmetry built in.
\section{Lattice QCD}
QCD most likely exists as a field theory due to asymptotic
freedom, in contrast to other field theories like QED, which can
only be viewed as effective models, and loose their validity at
short distances. In rigorous terms this means that the euclidean
Feynman integral which defines the theory
\begin{equation}
Z = \int\left[\prod_{\mu,x} d A_\mu(x)\right] \exp(-\beta S)\qquad
\beta = \frac{2 N_c}{g^2}
\label{eq:e3}\end{equation}
admits a thermodynamical limit.

$Z$ is a functional integral, and is defined as the limit of a
sequence of discretized, ordinary integrals. Lattice is an
approximant in this sequence, and a good approximation to the
limit whenever the correlation length is large compared to lattice
spacing, and small compared to the lattice size\cite{wil}.

Physical quantities computed on the lattice, e.g. by numerical
simulations, are therefore determined from first principles.
Lattice formulation also defines the correct ground state
(vacuum). The usual perturbative quantization in terms of Fock
space of gluons and quarks is instead basically ill defined, and
the instability of its ground state manifests itself by the
existence of singularities (renormalons) in the resummation of the
perturbative series\cite{3}. Lattice is therefore specially suited
to understand the structure of the theory.

A deep issue in this understanding is the $N_c\to \infty$ limit.
The idea\cite{2} is that the limit $N_c\to \infty$ of QCD, with
$\lambda = g^2 N_c$ fixed, defines a field theory, which contains
the main physical features of QCD, including confinement of
colour. An expansion in $1/N_c$, $N_f/N_c^2\ldots$ is well defined
and is a correction to the limiting theory. A fermion loop in this
expansion is roughly speaking ${\cal O}(N_f/N_c^{V/2})$ where $V$
is the number of gluon-quark vertices. For $V>2$ this is a small
correction, and only the loops with $V=2$ are ${\cal O}(1)$. Quark
loops are negligible except for the rescaling of $\Lambda_{Lattt}$
due to contribution of quarks loop with $V=2$ to the $\beta$
function. Numerical simulations on lattice do support this
expectation.

An alternative evidence in this direction is the solution of the
$U(1)$ problem\cite{A}, where the mass of the $\eta'$, $m_{\eta'}$
is related by $1/N_c$ arguments to the topological susceptibility
$\chi$ of the quenched QCD vacuum\cite{B}
\begin{equation}
2 N_f \chi = f_\pi^2 m_{\eta'}^2 \label{eq:e4}\end{equation}
where
\begin{equation}
\chi = \int d^4x \langle
T(Q(x)Q(0))|0\rangle\label{eq:e5}\end{equation} $Q(x)$ is the
topological charge density, $N_f$ the number of light fermions,
$f_\pi$ the pion decay constant.

Eq.(\ref{eq:e4}) predicts $\chi (180\,{\rm MeV})^4$ and this
prediction is confirmed by lattice simulations\cite{C}.

All this points to the independence of the confinement mechanism on $N_c$,
$N_f$.

The change of symmetry leading to confinement should not depend on $N_c$, $N_f$.
\section{Finite temperature QCD}
The standard way to define a static thermodynamics of QCD is to limit the
euclidean time direction from 0 to $t_E$
\begin{equation} t_E = 1/T \label{eq:e6}\end{equation}
with periodic boundary conditions for bosons, antiperiodic for fermions.

In lattice QCD an asymmetric lattice is used $N_s^3\times N_t$, with $N_s \gg
N_t$. At a given value of the lattice unrenormalized coupling $\beta = 2
N_c/g^2$ the temperature will be defined by eq.(\ref{eq:e6}) as
\begin{equation}
a(\beta) N_t = \frac{1}{T} \label{eq:e7}\end{equation}
in terms of the lattice spacing $a(\beta)$.

The value of $a(\beta)$ in physical units, depends on $\beta$ via
renormalization group. At one loop
\begin{equation}
a(\beta) = \frac{1}{\Lambda_{Latt}} \exp(-b_0\beta)\label{eq:e8}\end{equation}
where $b_0 > 0$ is the negative of the   first coefficient of the perturbative
expansion of the $\beta$ function.

At a given $\beta$ then
\begin{equation}
T = \frac{\Lambda_{Latt}}{N_t}
\exp(b_0\beta)\label{eq:e9}\end{equation} Temperature increases
exponentially with $\beta$: the deconfining transition from
hadrons to quark gluon plasma will take place at some value of
$\beta$. For $T > T_c$, or in the weak coupling regime, the order
parameter of the theory, the Polyakov line $\langle P \rangle$ is
$\neq 0$, and quarks and gluons exist as particles\cite{D}. For $T
< T_c$, $\langle P \rangle = 0$, and there is confinement. In
general $\langle P \rangle$ is related to the chemical potential
of a single quark, $\gamma$, by the equation
\begin{equation}
\langle P \rangle = \exp(-\gamma)\label{eq:e10}\end{equation}
so that the confined phase corresponds to $\gamma = \infty$: an infinite amount
of energy is needed to create an isolated quark. From the point of view of
symmetry $T < T_c$ is the disordered phase, or the strong coupling regime.

If confinement is due to symmetry, the obvious question is: what is the
symmetry of a disordered phase? As we shall see the key word in this respect is
``duality''.

Before going to duality we underline two points:
\begin{itemize}
\item[a)] $\langle P \rangle$ is an order parameter only in the absence of
quarks. In view of the $N_c\to\infty$ argument presented above,
such situation is not satisfactory. A order parameter should exist
for the two situations.
\item[b)] Evidence from lattice exists that confinement takes
place in QCD. Below $T_c$ the $Q \bar Q$ potential $V(R)$, as
determined from Wilson loop, is\cite{7}
\begin{equation}
V(R) = \sigma R\qquad \sigma = {\hbox{string tension}}
\label{eq:e11}\end{equation} Above $T_c$ the force is a screened
Coulomb force. Moreover in the confined phase the linear potential
is related to the existence of chromoelectric flux tubes joining
the $Q \bar Q$ pair, whose energy is proportional to the
length\cite{8,9}
\end{itemize}
\section{Duality}
Duality is a typical property of ($d+1$) dimensional systems which can have
spatial ($d$ dimensional) configurations with non-trivial topology and
conserved topological charge. At low $g^2$ the system is described in terms of
local fields $\Phi$, the symmetry is identified in terms of order
parameters $\langle\Phi\rangle$ and there is order.
 At high $g^2$, $\langle\Phi\rangle\to 0$, and disorder sets in.

However a dual description exists, in which the extended structures with
topology are described by local fields, $\mu$, the original $\Phi$ fields
become non local, and the effective coupling constant is $g'\sim 1/g^2$. In
this description the original disordered phase is ordered and viceversa.
The dual symmetry is described by $\langle\mu\rangle$ which is called a disorder
parameter.

The prototype model of duality is the 2 dimensional Ising model,
where the topological configurations are kinks\cite{18}.

Other examples are liquid $He_4$, with vortices\cite{19}, the
Heisenberg ferromagnet\cite{19a}, with non abelian vortices or
Weiss domains, and the usual 3+1 dimensional gauge theories, with
monopoles. $U(1)$ is an example\cite{19b}, SUSY QCD\cite{SW} is
another and finally QCD, as we shall show in what follows.

The strategy that we adopt is to identify the dual symmetry, and
to construct the disorder parameter $\langle\mu\rangle$ in terms
of the original local fields $\Phi$, which in QCD are gluons and
quarks. This strategy is inspired by ref.\cite{20} and has been
adapted on lattice by our group. An alternative strategy consist
in performing the transformation to dual\cite{18,E}, but is less
convenient in numerical investigations.
\section{Monopoles in QCD}
A guess of the dual symmetry in QCD is that the vacuum in the
confined phase is a dual superconductor\cite{5}. The
phenomenological motivation is that in a dual superconductor dual
Meissner effect will constrain the chromoelectric field acting
between a $Q \bar Q$ pair into an Abrikosov like flux tube, with
energy proportional to the length, or to the distance
\begin{equation}
V(R) = \sigma R \label{eq:e12}\end{equation} Monopoles are objects
with nontrivial topology, when described in terms of the gauge
fields. The signal of dual superconductivity should be the
nonvanishing of the vacuum expectation value of an operator
carrying magnetic charge, the analog of the Ginsburg Landau
charged Higgs field in ordinary superconductivity\cite{F}.

Monopoles in QCD are well understood\cite{14,15}. They can be
exposed by a procedure known as abelian projection\cite{16}. We
shall sketch it for SU(2) gauge fields: the extension to SU(N) is
trivial.

To each operator $\Phi(x) \equiv \vec\Phi(x)\cdot\frac{\vec\sigma}{2}$ in the
adjoint representation a gauge transformation $R(x)$ can be associated, which
brings $\vec\Phi$ parallel to 3 axis, or $\Phi(x)$ to a diagonal $2\times2$
matrix, proportional to $\sigma_3$. $R(x)$ is called ``abelian projection'' and
is defined everywhere except at zeros of $\vec\Phi(x)$. We shall define for
convenience
\begin{equation}
\hat\Phi(x) =
\frac{\vec\Phi(x)}{|\vec\Phi(x)}\label{eq:e13}\end{equation} The
field strength\cite{14}
\begin{equation}
F_{\mu\nu} = \hat\Phi\cdot\vec G_{\mu\nu} -
\frac{1}{g}\hat\Phi\cdot( D_\mu\hat\Phi \wedge D_\nu\hat\Phi)
\label{eq:e14}\end{equation} with $\vec G_{\mu\nu}$ the usual
field strength and $D_\mu = \partial_\mu - g \vec A_\mu\wedge$ the
covariant derivative, is a gauge invariant, colour singlet
operator. By construction the bilinear term in $A_\mu$ cancels
between the two terms in eq.(\ref{eq:e14}) and in the abelian
projected gauge
\begin{equation}
F_{\mu\nu} = \partial_\mu A^3_\nu - \partial_\nu A^3_\mu
\label{eq:e15}\end{equation}
looks like an abelian field. Its dual $F^*_{\mu\nu} =
\frac{1}{2} \varepsilon_{\mu\nu\rho\sigma} F^{\rho\sigma}$ defines a magnetic
current $j^M_\mu$ as
\begin{equation}
j^M_\mu = \partial_\nu F^*_{\mu\nu} \label{eq:e16}\end{equation}
which is identically conserved
\begin{equation}
\partial_\mu j^M_\mu = 0 \label{eq:e17}\end{equation}
A magnetic, gauge invariant $U(1)$ symmetry is thus associated to each operator
$\Phi(x)$ belonging to the adjoint representation.

$j^M_\mu$ is not trivially zero: monopoles charges can exist at
the sites where $\vec\Phi(x)=0$, where the abelian projection
$R(x)$ is singular\cite{16}.

Dual superconductivity will occur when magnetic charges belonging
to the above $U(1)$ condense in the vacuum, and the corresponding
electric field, i.e. the chromoelectric field parallel to the 3
color axis after abelian projection, will experience dual Meissner
effect.

The disorder parameter $\langle\mu\rangle$ describing monopole
condensation can be constructed with a technique tested in a
variety of known system\cite{19,19a,19b}, following the strategy
outlined above. We will not go into the details of the
construction, for which we refer to\cite{17}.

The transition to superconductor is signalled by a sharp negative
peak in the plot of the quantity $\rho = \frac{d}{d\beta}
\ln\langle\mu\rangle$ versus $\beta$ (or vs T). The peak becomes
sharper as the space volume of the system becomes larger. A
typical behaviour is shown in fig.(1).
\vskip0.1in
\noindent\hskip0.2\textwidth
\begin{minipage}{0.7\textwidth}
\epsfig{figure=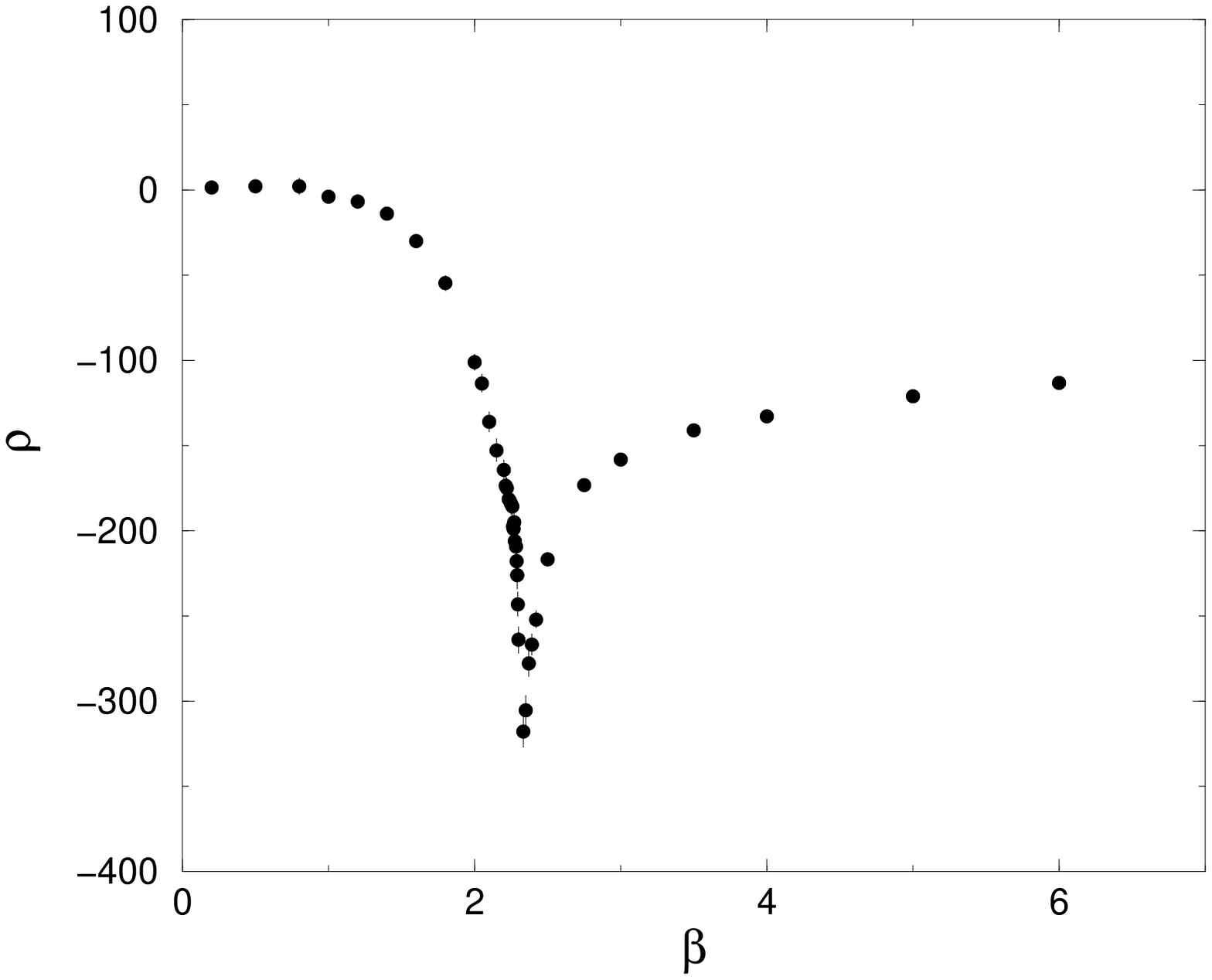, width=7.8cm} \par\noindent Figure 1:
{$\langle \rho \rangle$ vs. $\beta$ for $SU(2)$ gauge theory.\\
Plaquette projection, lattice $16^3 \times 4$.}
\end{minipage}
\vskip0.1in
 A finite size scaling analysis allows to extract the critical
indices and the decoupling temperature\cite{17,19,19a,19b}.
Indeed, for dimensional reasons
\begin{equation}
\mu =
L_s^d\Phi(\frac{a}{\lambda},\frac{\lambda}{N_s})\end{equation}
where $a$ is the lattice spacing, $\lambda$ is the correlation
length, $N_s$ the extension of the system. At $\beta_c$, $\lambda$
diverges as $\lambda\simeq(\beta_c-\beta)^{-\nu}$, with $\nu$ the
correlation critical index, and $a/\lambda \ll 1$. In the limit
$a/\lambda\to 0$ the scaling law follows for $\rho =
\frac{d}{d\beta} \ln\langle\mu\rangle$
\begin{equation}
\frac{\rho}{L_s^{1/\nu}} = \Phi(L_s^{1/\nu}(\beta_c - \beta))
\end{equation}
The quality of scaling is shown in fig.2 for $SU(2)$ and in fig.3
for $SU(3)$. Scaling exists only for the proper values of
$\beta_c$, $\nu$, which can thus be determined.
\setcounter{figure}{1}
\noindent
\begin{figure}
\begin{minipage}[htb]{7.8cm}
\centering{\epsfig{figure=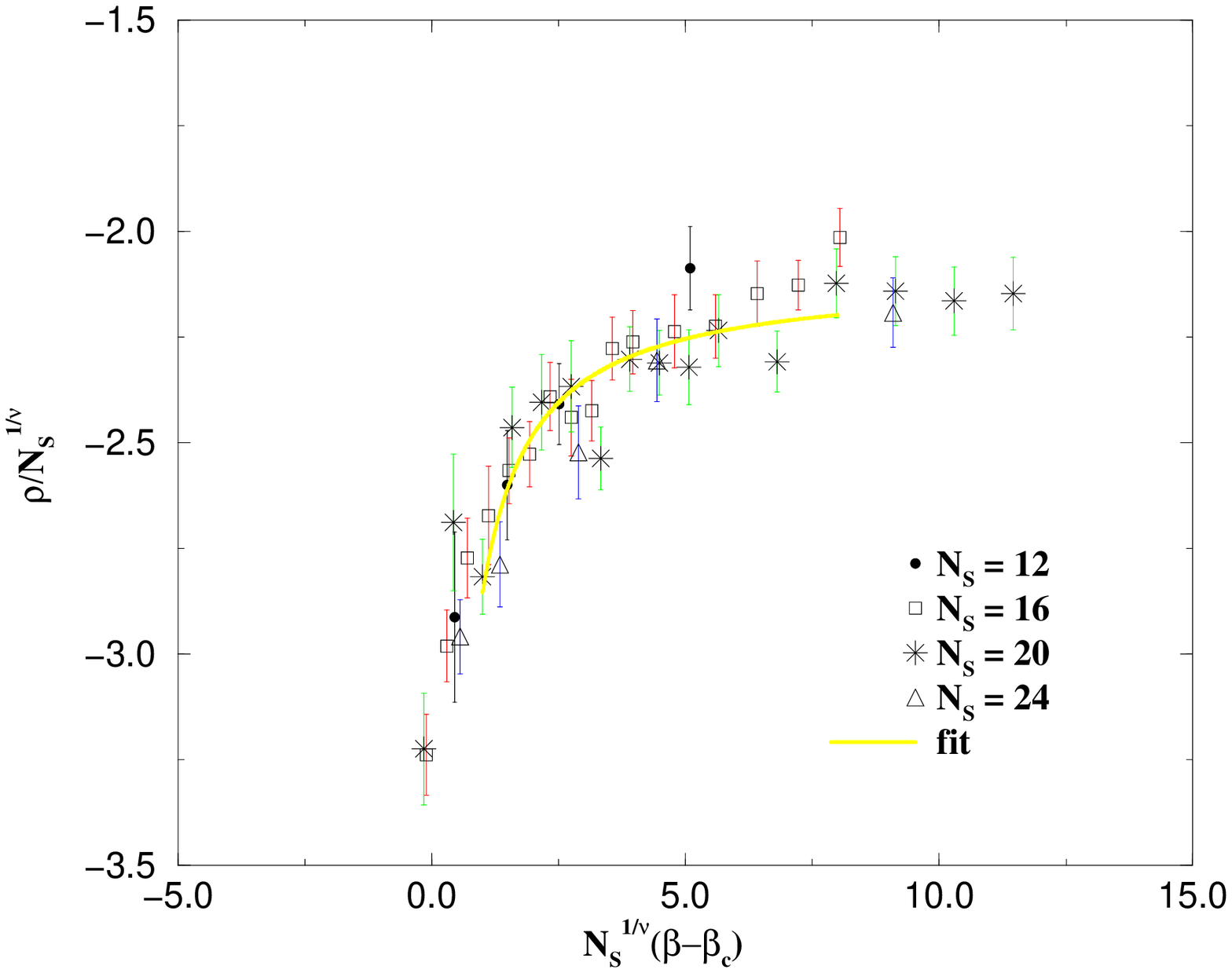, width=8cm}}
\caption{Quality of the scaling for $SU(2)$ gauge theory.
Plaquette projection, $N_t = 4$.}
\label{fig5}
\end{minipage}
\hfill
\begin{minipage}[htb]{7.8cm}
\centering{\epsfig{figure=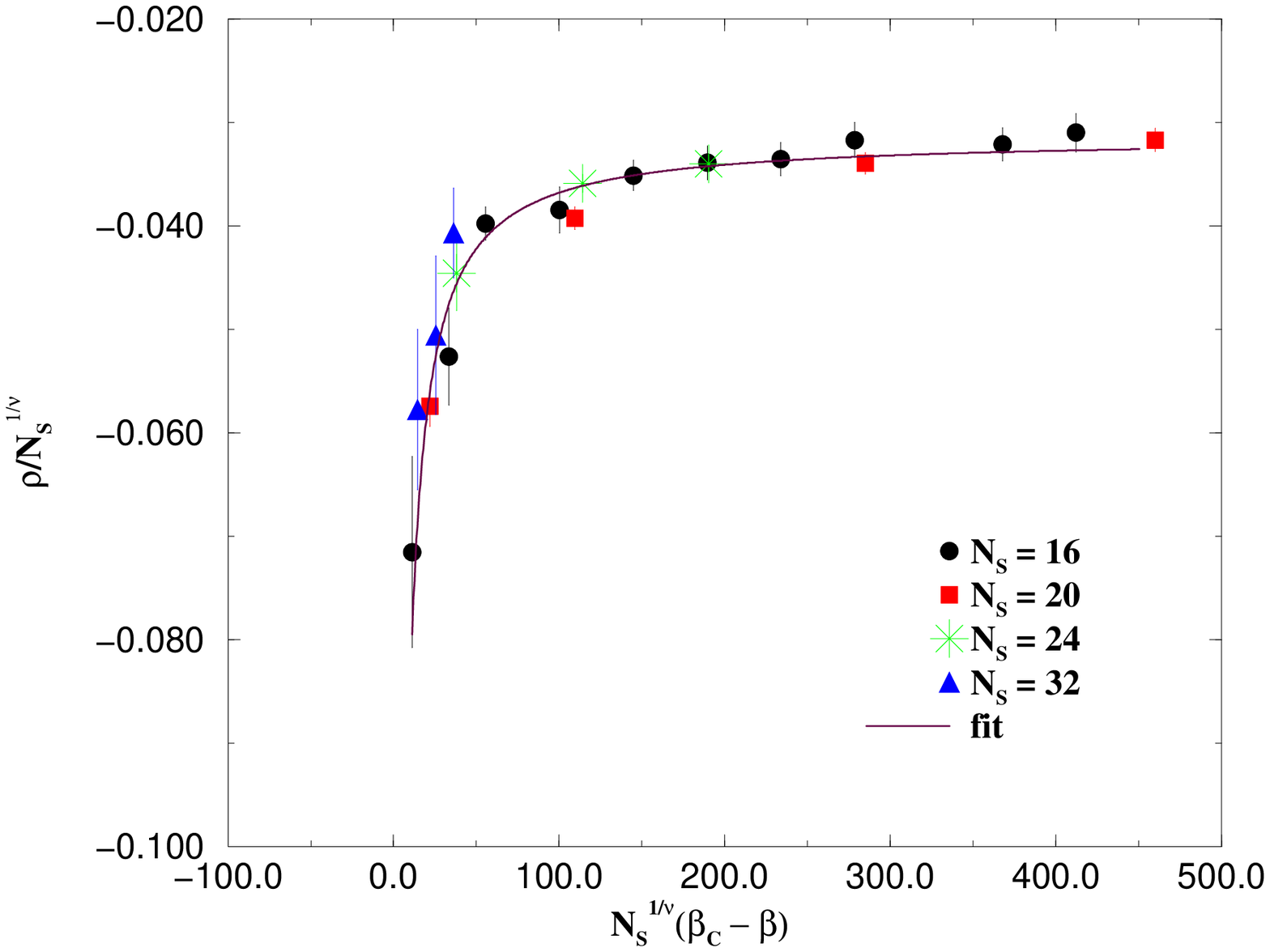, width=7.8cm}}
\vspace{0.2cm}
\caption{Quality of the scaling for $SU(3)$ gauge theory.
Polyakov projection, $N_t = 4$.}
\label{fig6}
\end{minipage}
\end{figure}

These quantities can also be determined by different
methods\cite{D}. They agree within errors. For $SU(2)$ we find
\begin{eqnarray}
\nonumber
\hspace{4cm}
\begin{array}{l}
\nu = 0.63(5)\\
\beta_c = 2.30(2) \qquad N_t = 4 \\
\delta = 0.20(8)
\end{array}
\ .
\end{eqnarray}
For $SU(3)$
\begin{eqnarray}
\nonumber
\hspace{4cm}
\begin{array}{l}
\nu = 0.33(2)\\
\beta_c = 5.70(3) \qquad N_t = 4 \\
\delta = 0.54(4)
\end{array}
\ ,
\end{eqnarray}

\section{Discussion: open problems.}
There is definite evidence that confinement is produced by dual
superconductivity. What has not been specified above is the field
$\vec\Phi(x)$. In principle monopoles exist for any choice of it:
a functional infinity of monopoles species. There is no argument a
priori that all of them should condense. In ref\cite{16} t'Hooft
guesses that all monopoles are physically equivalent, and all of
them should condense in the confined phase. What we observe on
lattice supports this view. A few choices for $\Phi$ that we have
tried look perfectly equivalent.

An alternative attitude exists in the literature, and is that the
so called ``maximal abelian projection'' defines monopoles which
are more equal than others. In fact with that choice the abelian
field dominates, since after the projection all the fields are
oriented in its direction to $90\%$. This choice can be more
convenient than others if one looks for effective lagrangians. In
principle this convenience does not put any limitation to the
symmetry pattern of confinement.

Most probably a better and more syntetic understanding of this
symmetry is needed. There must be a reason why all abelian
projections are equivalent. However there is no doubt that dual
superconductivity is the mechanism of colour confinement.

\vspace{0.5cm}
The part of the work reported due to our group has been done in collaboration
with L. Del Debbio, G. Paffuti, P. Pieri, B. Lucini, D. Martelli in the last
few years. Their contribution was determinant to the results.

\end{document}